\newcommand{\aap}{{\it Astron. Astrophys.}}

\newcommand{\apjl}{{\it Astrophys. J. Lett.}}
\newcommand{\apjs}{{\it Astrophys. J. Supp. Ser.}}
\newcommand{\apss}{{\it Astrophys. Space Sci.}}

\newcommand{\mnras}{{\it Mon. Not. R. Astron. Soc.}}
\newcommand{\nat}{{\it Nature}}
\newcommand{\na}{{\it New Astronomy}}

\documentclass{iau_FM}
\usepackage{amsmath}
\usepackage{graphicx}
\usepackage{hyperref}

\title[PASO] 
{PASO - Astronomy and Space Situational Awareness in a Dark Sky Destination}

\author[Domingos Barbosa et al.]   
{Domingos Barbosa$^1$
 , Bruno Coelho$^{2,5}$, Miguel Bergano$^{2,4}$, Constança Alves$^{1,3}$, Alexandre C.M. Correia$^{5,8}$, Lu\'{\i}s Cupido$^{12}$, José Freitas$^9$, Lu\'{\i}s Gon\c{c}alves$^{1,5}$, Bruce Grossan$^{14}$, Anna Guerman$^7$, Allan K. de Almeida Jr$^{2,4}$, Dalmiro Maia$^4$, Bruno Morgado$^4$, João Pandeirada$^{1,6}$, Val\'erio Ribeiro$^{10}$,  Gon\c{c}alo Rosa$^{11}$, George Smoot$^{14,15}$, Timoth\'ee Vaillant$^{2,4,5}$, Thyrso Villela$^{13}$, Carlos Alexandre Wuensche$^{13}$}

\affiliation{$^1$Instituto de Telecomunica\c{c}\~oes, Universidade de Aveiro, \\Campus Universitário de Santiago, 3810-193, Aveiro, Portugal  \\ email: {\tt dbarbosa@av.it.pt}\\[\affilskip]
$^2$ ATLAR, Ed. Multiusos, Rua Rangel de Lima, 3320-229 Pampilhosa da Serra, Portugal\\ [\affilskip]
$^3$University of Aveiro, Campus Universitário de Santiago, 3810-193, Aveiro, Portugal  \\[\affilskip]
$^4$ CICGE, Faculdade de Ci\^encias da Universidade do Porto, 4169-007 Porto, Portugal\\ [\affilskip]
$^5$ CFisUC, Departamento de F\'{\i}sica, Universidade de Coimbra, 3004-516 Coimbra, Portugal\\ [\affilskip]
$^6$ Instituto Superior T\'ecnico, Avenida Rovisco Pais 1, 1049-001 Lisboa, Portugal\\[\affilskip]
$^7$ C-MAST, Center for Mechanical and Aerospace Science and Technology, \\University Beira Interior, 6201-001 Covilh\~a, Portugal\\[\affilskip]
$^8$ IMCCE, UMR8028 CNRS, Observatoire de Paris, PSL Universit\'e, \\77 Avenue Denfert-Rochereau, 75014
Paris, France \\[\affilskip]
$^9$ Dep. of Space, Armed Forces General Staff, Avenida Ilha da Madeira, \\1449-004 Lisboa, Portugal \\[\affilskip]
$^{10}$ KLM, Data \& Technology - Platform Ground, \\PO Box 7700, 1117 ZL Schiphol, the Netherlands\\ [\affilskip] 
$^{11}$ Ministry of Defense, Avenida Ilha da Madeira, 1449-004 Lisboa, Portugal \\ [\affilskip]
$^{12}$ LC Technologies, Rua da Vila Verde 39, 3800-810 Eixo / Aveiro, Portugal \\
$^{13}$  Instituto Nacional de Pesquisas Espaciais, Divis\~ao de Astrof\'{\i}sica, Av. dos Astronautas, 1758, 12227-010 - São Jos\'e dos Campos, SP, Brazil\\ [\affilskip]
$^{14}$ Lawrence Berkeley National Laboratory, 1 Cyclotron Rd, Berkeley, CA 94720, United States\\ [\affilskip]
$^{15}$ University of California, 366 Physics North, MC 7300, Berkeley, CA 94720-7300, United States

}

\pubyear{2023}
\volume{S385}  
\pagerange{119--126}
\jname{Astronomy and Satellite Constellations: Pathways Forward}

\editors{C. Walker,  D.Turnshek, P.Grimley, D.Galadi-Enriquez \&  M.Aub\'e}
\begin{document}

\maketitle

\begin{abstract}
The Pampilhosa da Serra Space Observatory (PASO) is located in the center of the continental Portuguese territory, in the heart of a certified Dark Sky destination by the Starlight Foundation  (Aldeias do Xisto) and has been an instrumental asset to advance science, education and astrotourism certifications. PASO hosts astronomy and Space Situational Awareness (SSA) activities including a node of the Portuguese Space Surveillance \& Tracking (SST) infrastructure network, such as a space radar currently in test phase using GEM radiotelescope, a double Wide Field of View Telescope system, a EUSST optical sensor telescope. These instruments  allow surveillance of satellite and space debris in LEO,  MEO and GEO orbits. The WFOV telescope offers spectroscopy capabilities enabling light curve analysis and cosmic sources monitoring. Instruments for Space Weather are being considered for installation to monitor solar activities and expand the range of SSA services. 
\end{abstract}
\firstsection 
\section{Early days: from GEM to SSA}

The number of space activities, including the launch of constellations of satellites and the associated space debris population, has been steadily increasing, raising concerns over the impact on the night sky quality for astronomical and space science purposes.  Astronomy and Situational Awareness services have been developed at PASO, in Pampilhosa da Serra municipality in the Central Region in Portugal since 2011. This space observatory has been instrumental to foster astronomy and space science, citizen science and astrotourism contributing to showcase the sky quality of the Dark Sky "Aldeias do Xisto" ("Schale villages") tourist destination at Pampilhosa da Serra territory. This Dark sky destination is named after its stone (shale) widely used in traditional construction across the region.

PASO took its origin after a site selection to install a radiotelescope for the Galactic Emission Mapping project in Portugal (GEM-P), \cite{2006NewA...11..551F}, with a careful site survey on clima, low Radio frequency Interference (RFI) levels - location should be a radio quiet site - and infrastructure capabilities. The GEM project was an international collaborative effort in the fields of radio astronomy and cosmology. The project goal was to accurately determine the spatial distribution and absolute intensity in the radio and microwave spectrum of the large scale distribution of the synchrotron radiation emitted by the Milky Way galaxy and by the unresolved blend of external galaxies, see \cite{1996Ap&SS.240..225T}. 


The GEM experiment concept was pioneered by the Berkeley team to address the problems of CMB foreground cartography at the time of the detection of temperature anisotropies in the CMB radiation by the Differential Microwave Radiometers (DMR) onboard the COBE satellite \cite{1992ApJ...396L...1S}, a major scientific breakthrough for cosmology. COBE DMR all-sky maps at 30 GHz, 53 GHz, and 90 GHz have shown the need to extend our knowledge of diffuse Galactic emissions, \cite{1992ApJ...396L...7B}, and eliminate foreground contamination from pristine cosmic signals. Since COBE,  high resolution map-making made a huge leap forward from the data returned from more recent experiments lie, MAXIMA,\cite{2000ApJ...545L...5H,2001ApJ...561L...7S}, Boomerang, \cite{2000Natur.404..955D}, DASI, \cite{2002Natur.420..763L}, NASA’s WMAP, \cite{2013ApJS..208...20B}, and ESA’s Planck Surveyor satellite, \cite{2010A&A...520A...1T, 2020A&A...641A...6P},  that offer a direct glimpse into the physics at the cosmic surface of last scattering, providing constraints on cosmological parameters, tests of theories of large scale structure formation and favoring the inflationary paradigm \footnote{For a comprehensive list of CMB and Foreground experiments: NASA Legacy Archive for Microwave Background Data Analysis (LAMBDA); \url{https://lambda.gsfc.nasa.gov/}}.

\begin{table}\label{equipments}
  \begin{center}
  \caption{PASO SSA instruments }
  \label{tab2}
 {\scriptsize
  \begin{tabular}{|l|c|c|c|c|c|c| }\hline 
{\bf Radio Instruments} & {\bf Frequency} & {\bf Size} & {\bf Mount} & {\bf Type} & {\bf Speed} & {\bf FoV}$^3$  \\
     &           &     &    &   &  &  \\ \hline
L-Band radiotelesescope      & $~1.42$GHz   & $~5$-m  & Alt-Azi & Front-fed parabolic &  & $2.9^o$ \\ 
(Science, Outreach)          &             &           &        &                     & &  \\
C-Band space radar$^1$ & $~5.56$GHz      &  $9$-m               & Alt-Az& Cassegrain & $\le 10^o$/sec & $0.5^o$ \\ 
(LEO  ($< 1000$ Km)  )        &             &           &        &                     & &   \\

Solar Radiospectrograph$^2$ & $0.1$-$1$GHz & $5$-m             & Alt-Azi & Front-fed parabolic &  & $2^o$ (2GHz) \\ 
(Space Weather)          &             &           &        &                     & &   \\ \hline
{\bf Optical Instruments}      &  {\bf Bands}   & {\bf Size}   &  {\bf Mount}   &  {\bf Type} & {\bf Speed} & {\bf FoV} \\
                &           &   &      &   &  & \\ \hline

Twin WFOV Telescope$^4$  &   BVRI,H$\alpha$    &       2x0.3 m             &  Equ & Reflector Cassegrain &  $\le 40^o$/sec & $4.3^o$ x $2.3^o$  \\
  (LEO - GEO)              &      OIII     &   &      &  &  &  \\ 
WFOV Optical Telescope$^5$  & I                      & $0.28$ m       &  Equ  & RASA 11"& $\le 6^o$/sec & $\sim 2.6$x$2.6^o$ \\ 
      (MEO/GEO)          &           &   &      &   & &  \\ \hline


  \end{tabular}
  }
 \end{center}
\vspace{1mm}
 \scriptsize{
 {\it Notes:}\\
  $^1$ Former GEM radiotelescope\\
  $^2$ In Commissioning by FCUP. Expected start of operations for mid-2024, pending protocol and final site agreement with PT MoD. \\
  $^3$ FoV is taken as beamwidth at -3dB.\\
  $^4$ It-UC-MDN partnership.\\
  $^5$ EUSST MDN telescope.}
\end{table}

GEM was first designed as a portable and double-shielded 5.5-m radiotelescope, a constantly rotating ground platform in altazimuthal configuration to map 60-degree-wide declination bands from different observational sites with a slow circular scan pattern of the sky at zenithal angles of 30°. This GEM design characteristic proved to be of great interest for Space Surveillance \& Tracking of satellites and space debris, as is discussed in chapter \ref{sst}. GEM made observations from several locations, from Bishop, California, Leyva in Colombia which is close to the equator, Teide, Tenerife, Spain and settled in Cachoeira Paulista, Brazil, to maximize homogeneous sky observations. 


Since the beginning of this project, total power maps of the diffuse galactic synchrotron in 408, 1465 e 2300 MHz were accomplished with total power receivers, with later addition of custom developed 5GHz polarimeters to provide I,Q and U Stokes parameter maps, see \cite{2011ExA....30...23B, ivan2008}. These frequencies were important to cover the frequency gap at centimetre wavelengths on synchrotron radiation. Meanwhile, other projects like C-BASS, \cite{2022MNRAS.513.5900H}, followed and embraced similar concepts to improve our knowledge of the diffuse galactic emissions.
To integrate for large sky areas and since sensitivity is more important than resolution, GEM scanning strategy consists on an azimuthal dish rotation until the required sensitivity is attained, thus allowing integration for large sky areas and provided useful foreground templates to CMB teams. The Moon was used to calibrate the antenna temperature scale and the preparation of the map required direct subtraction and destriping algorithms to remove ground contamination as the most significant source of systematic error, \cite{2013A&A...556A...1T}. To add the northern hemisphere coverage, and thus produce a template of most of the sky, Portugal's PASO location was assessed as a suitable location to host a 9-metre antenna to complement the original GEM southern hemisphere maps and enable future expansion towards observation of polarized galactic diffuse emission. The GEM working frequencies were chosen because below 20 GHz, atmosphere contribution and polarization contamination are negligible. Improved mapping of the spectral and large-scale properties of Galactic synchrotron radiation would result from an observational strategy aimed to minimize ground contamination and long-term systematics to guarantee baseline uniformity. However, we acknowledge concerns on the spectral occupancy of satellite communications in C and X bands could severely impact the map-making observation of future sensitive galactic surveys.

 \begin{figure}
    \centering
    \includegraphics[width=0.9\linewidth]{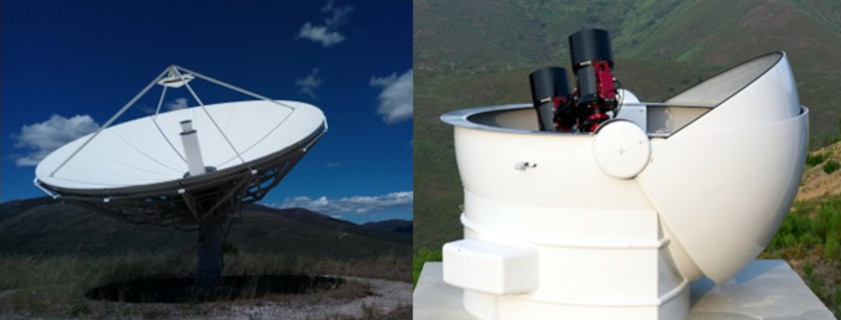}
    \caption{Left: ATLAS, the PASO 5.56 GHz monostatic radar tracking sensor; Right: the PASO WFOV Double telescope System.}
    \label{fig:teleleo}
\end{figure}

\section{Space Surveillance \& Tracking} \label{sst}

\begin{table}
  \begin{center}
  \caption{Overview of current satellite Constellations major parameters
}
  \label{tab1}
 {\scriptsize
  \begin{tabular}{|l|c|c|c|}\hline 
{\bf Parameter} & {\bf GEO} & {\bf MEO} & {LEO}  \\
                &           &         &    \\ \hline
{\bf Altitude}      & $~36,000$ Km & $~5,000$ -  $20,000 $ Km & $~500$ - $1,200$ Km \\ 
                    &              &                          & \\ \hline
{\bf Coverage area} & Vast      & Medium               & Narrow \\ 
                    &           &                      &  \\ \hline
{\bf Downlink and uplink}  & Slow & Medium             & Fast \\ 
{\bf rate (signal speed)}  &      &                    & \\\hline
{\bf Ground station}       & Distant & Regional        & Local  \\
{\bf spacing}              &         &                 & \\ \hline

  \end{tabular}
  }
 \end{center}
\vspace{1mm}
\end{table}

The GEM-P team used a 9-metre diameter Cassegrain Vertex antenna with alt-az mounting. This antenna was originally a telecommunications antenna for C-band operations with a King Post pedestal mounting. The new observational strategy and mechanical modifications  allowed a steady, constant, slow azimuthal rotation of 0.3-1rpm at fixed elevation of 30º angle from zenith, while allowing for travel speeds up to 6 deg/sec in azimuth. The operational requirements for the GEM-P antenna with precise pointing at constant speed implied some key mechanical and structural changes and enabled actually novel observation strategies for space debris surveillance, \cite{pandeirada2024}. Further refurbishments meant the antenna is able to cope with fast angular speeds and therefore responds to operational requirements of LEOs surveillance. Therefore, the equipment became included in the Portuguese node of the European Space Surveillance \& Tracking program (EUSST)\footnote{https://www.eusst.eu/} under contract with the Portuguese Ministry of Defense (PT MoD). The system operates at 5.56 GHz, with a beamwidth of 0.73$^o$ and aims to provide range and range-rate measurements of objects in LEO with RCS above 10 cm$^2$ at 1000 Km (Table \ref{equipments}).

Additionally, PASO got the installation of a double wide field of view telescope system with a maximum FoV of 4.3$^o$ x 2.3$^o$ (see Fig. \ref{fig:teleleo}). The sensor consists of two small aperture (30cm) telescopes. The system can observe in white light, in BVRI bands, and in the narrow bands H(alpha) and O[III]. The equatorial mount has very fast speeds (max. slewing speed of 40$^o$/s) enabling tracking capability of LEO objects. This design is meant to support science and space surveillance services with its own scheduling tools. Besides EUSST, the WFOV system was designed to enable observations of transient sources and provide regular monitoring of novae and other powerful cosmic nuclear explosions like novae \cite{2021ApJS..257...49C} and contribute in the near future to surveillance networks of gravitational-wave sources like GRANDMA, \cite{2022SPIE12186E..1HA} pending a Concept of Operations (CONOPS) final agreements. PASO location has excellent all sky clearance, in a hilltop at 840 m, and surrounded by mountain ranges that allow protection from the light pollution of the coastal regions of the country. The conditions are among the best in continental Portugal for optical use in SST operations, where sky background reaches magnitude 21 or above with more than 200 clear nights per year (on average) \cite{2021arXiv210702315C}.

Near-future sensor fusion concepts using space radar and very wide-field tracking telescopes for Space Surveillance and Tracking (SST), such as the PASO telescopes, \cite{2022arXiv221104443C}, will provide additional information that is much needed to accurately model the orbital parameters and attitudes of satellites and debris, \cite{2024AcAau.215..548D}.

Using several types of sensors in the same place as PASO has many benefits:  radiotelescope and associated space radar measurements give precise radial velocity and distance to the objects, while the optical telescopes gives better sky coordinates measurements. With the installation of radar and optical sensors, PASO can extend the observation time of space mega constellations and space debris and correlate information from optical and radar provenances in real time. During twilight periods both sensors can be used simultaneously to rapidly provide orbital parameters and  enable computing of new TLEs for LEO objects, eliminating the time delays involved in the data exchange between sites in a large SSA/SST network. This concept will not replace the need for a SST network with sensors in multiple locations around the globe, but will provide a more complete set of measurements on object passages and therefore increases the added value for initial orbit determination or monitoring of atmospheric reentry campaigns. 

\subsection{Space Weather}

PASO is also a suitable location to address space  weather observations. Candidate projects include solar observation telescopes dedicated to monitoring and outreach in partnership with the local Pampilhosa da Serra Town Hall and the Faculty of Sciences of University of Porto (FCUP) solar radiospectrograph for radio observations (Table \ref{equipments}).   This solar radio instrument is a fully refurbished and upgraded  radiospectrograph, an instrument dedicated to monitor the strong radio emissions produced during the development of strong Space Weather events that waits final clarification on site commissioning due to
happen by mid 2024. This fast instrument will sweep the entire band from 100MHz up to 1GHz,  with a  bandwidth of 2MHz and 5msec integration time.

\begin{figure}[b]
\begin{center}
 \includegraphics[width=4.3in]{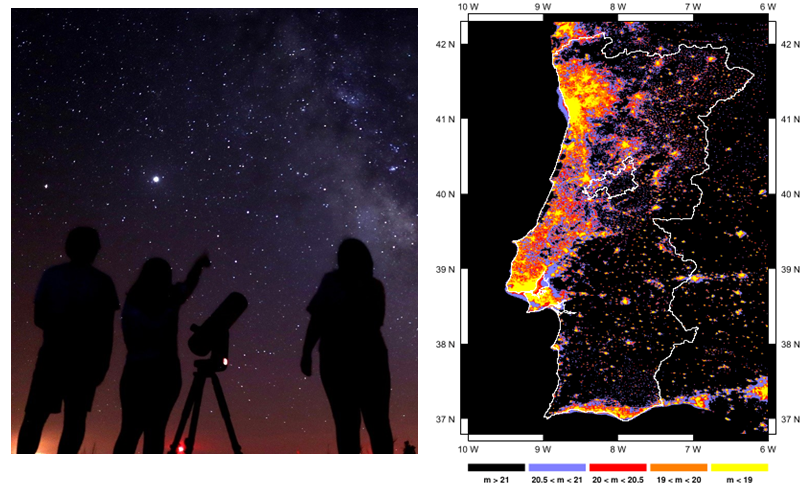} 
 \caption{Left: Cibercosmos activities with a new digital Unistellar eVscope: a summer Milky Way, from (\cite{2022AcAau.200..612B}); Right: Map of the magnitude in the V band for the Portuguese mainland. The contours of the area defined by the seven councils of reference are superposed on the map, from (\cite{dark_sim2018}).}
   \label{figdark}
\end{center}
\end{figure}
An earlier version of the FCUP radiospectrograph, \cite{maia_etal99}, had a large history of collaboration with  major solar instruments, namely with the Nançay Radioheliograph and also with the Large Angle and 
Spectrometric Coronagraph (LASCO) on board the Solar and Heliospheric  Observatory (SOHO)  on the detection and monitoring of several powerful and energetic solar events.

The refurbished and upgraded FCUP radiospectrograph will benefit from
new available instruments, in particular new radio imaging instruments
with great sensitivity and large spectral coverage like the SKA, 
  \cite{2015aska.confE.169N}. Potential contributions from the SKA
include interplanetary scintillation (IPS), radio-burst tracking, and
solar spectral radio imaging with a superior sensitivity. These will
provide new insights and results in topics of fundamental importance,
such as the physics of impulsive energy releases, magnetohydrodynamic 
oscillations and turbulence, the dynamics of post-eruptive processes, 
energetic particle acceleration, the structure of the solar wind and
the development and evolution of solar wind transients at distances up
to and beyond the orbit of the Earth.

The FCUP radiospectrograph design is adequate for the monitoring of a
wide range of various heliospheric phenomena of fundamental interest,
in particular relating to flare physics, Coronal mass ejection (CME)
initiation, shock propagation, and energetic particle  acceleration.
CMEs are recognized as primary drivers of disturbances in the
interplanetary medium. As such, they can have a profound impact on the
near-Earth environment. A long-standing problem has been understanding
the origin of CMEs and the details of their relationship to a number
of associated phenomena, including solar flares, coronal and
interplanetary shocks, and solar energetic particle (SEP) events. Radio
observations can be of paramount importance in capturing the formation
stages of the CME, and in following the CME evolution in the corona and
heliosphere, see \cite[Vourlidas \etal (2020)]{2020FrASS...7...43V}.

\section{Astrotourism}

Astrotourism and related citizen science activities are becoming a major trend of a sustainable, high-quality tourism segment, core elements to the protection of Dark skies in many countries. 
Indeed, dark skies are a natural asset that is becoming rarer due to light pollution at night. Therefore, it became recognized that preserving dark skies and developing a plethora of astronomical activities for the common citizen can be integrated into  sustainable and inclusive tourism experiences, contribute to the conservation of natural resources and cultural heritage. Moreover, astrotourism within dark skies locations requires minimal infrastructure and therefore presents the potential for good return on investment. As well described in \cite{2021ASPC..531...17D, Dunn Edensor}, the United Nations Societal Development Goals (SDGs) impacted by astrotourism and related science education activities for development through education, sustainable job creation, reduced inequalities, and the preservation of local tradition and indigenous knowledge impact at least SDGs 4, 5, 8, 9, 10, and 11: quality education; gender equality; decent work and economic growth; industry, innovation, and infrastructure; reduced inequalities; and sustainable communities. Good examples of this awareness were built through capacity-building projects such as the National Astrophysics and Space Science Programme \cite{2019NatAs...3..369B}, the Square Kilometre Array Human Capital Development Programme (HCDP), the DevelOpment of PaloP knowLEdge in Radioastronomy (DOPPLER) and the development in Africa with Radio Astronomy (DARA), \cite{2018NatAs...2..505H}.

Being an active observatory in expansion, PASO quiet and dark sky was thought as the epicentre of a new dark Sky destination. The GEM Portugal and FCUP teams started site testing campaigns and analysis of satellite data performed within the classical scheme of optical seeing properties, air transparency, sky darkness, and cloudiness. For the purpose of Starlight Tourist Destination certification those properties were checked to determine compliance with the following criteria:
\begin{itemize}
    \item Seeing better than 3 arcsec.
    \item A cloud free sky.
    \item Visual magnitude of the sky higher than 21 mag/arsec2.
    \item Sky transparency such that the Naked Eye LimitingMagnitude (NELM)
in the visible is higher than 6.
\end{itemize}

A methodology using the semi-empirical models presented in \cite{2001MNRAS.323...34C} inferred the area and the seven municipalities that fulfilled at least two of the two certification criteria. Using satellite observations we were able to infer the average values and time variability of cloud cover and aerosol content over the area of interest. By combining nighttime satellite imagery with field testing we were also able to infer the sky brightness in magnitudes in the area of interest. The analysis revealed that an area corresponding to most of the Lous\~a and Açor Mountains fulfills the criteria for certification regarding cloud cover, atmospheric transparency and sky brightness. The seeing measurements made as part of field work confirmed that the most promising locations are in the mountain ranges, at heights above 500 metres fulfilled the criteria for certification regarding cloud cover, atmospheric transparency and sky brightness. Finally, by 2018, an area encompassing 33 parishes in 14 municipalities was certified as part of the Dark Sky Tourist Destination "Aldeias do xisto" by the Starlight Foundation, endorsed by the UNWTO and UNESCO.

Local outreach activities boomed and in the Summer of 2020, in the middle of COVID pandemics, we started an initiative to train young students - Cyber-Cosmos - using an Unistellar eVscope, a smart, compact and user-friendly digital telescope that offers unprecedented opportunities for deep-sky observation and citizen science campaigns. Sponsored by the Ciência Viva Summer program, Pampilhosa da Serra, a region at the heart  of this certified Dark Sky destination, was the chosen location for this project where we expect astrotourism and citizen science to flourish and contribute to astronomy and space sciences education, \cite{2022AcAau.200..612B}. This was probably the first continuous application of this digital equipment in a pedagogical, citizen-science and pandemic context.

\section{Conclusions}

With its new capabilities, PASO science and space service programs like the EUSST can provide invaluable data to its agreed SSA activities and help understand the evolution of space megaconstellations and the space debris field orbiting our planet and contribute to astronomical science through the monitoring of transient cosmic sources. PASO constitutes a perfect site for the development and testing of new radar and optical data fusion algorithms and techniques for space debris monitoring and for  Dark and Quiet Sky quality regular assessment. From its early days, GEM Portugal paved the way for a plethora of scientific,  education, outreach and astronomy activities that flourished into a new observatory within a certified Dark Sky Destination. The local PASO SSA studies may be of future interest to contribute to the International Astronomical Union Centre for the Protection of the Dark and Quiet Sky from Satellite Constellation Interference (IAU CPS). Although concerns do exist with mounting evidence of the rise of light pollution trends that threatens this small dark and quiet spot in the the middle of Portugal, PASO contributed to public policy awareness to help protect the local night sky. 

\section*{Acknowledgements}
We would like to remember the memory of the late Ivan Ferreira, from INPE and UNB, Brazil. DB acknowledges partial support by the European Regional Development Fund (FEDER), through the Competitiveness and Internationalization Operational Programme (COMPETE 2020) of the Portugal 2020 framework [Project SmartGlow with Nr. 069733] (POCI-01-0247-FEDER-069733)]; IT, UC and FCUP acknowledge further support from ENGAGESKA Research Infrastructure, ref. POCI-01-0145-FEDER-022217, funded by COMPETE 2020 and Fundac\c{c}\~ao para a Ci\^encia e a Tecnologia (FCT), Portugal; IT team members acknowledge support from Projecto Lab. Associado UID/EEA/50008/2019. DM, BM, AKAJ and TV were partially supported by the project Centro de Investigação em Ciências Geo-Espaciais, reference UIDB/00190/2020, funded by COMPETE 2020 and FCT, Portugal. AG acknowledges support from Center for Mechanical and Aerospace Science and Technologies - C-MAST, funded by FCT through project UIDB/00151/2020. AC was supported by CFisUC (UIDB/04564/2020 and UIDP/04564/2020). JP acknowledges support by FCT through PhD grant No-2022-12341-BDANA. AKAJ acknowledges support from ATLAR Innovation and from Instituto Nacional de Pesquisas Espaciais (INPE)-Brazil. The team acknowledges support by the European Commission H2020 Programme under the grant agreement 2-3SST2018-20, additional funding from 1-SST2018-20 and the partnerships led my MoD and EMGFA. We are grateful for the EUSST / MoD support that enabled the planning, deployment and operation of new SSA instruments. The Team acknowledges support from the Town Hall of Pampilhosa da Serra and the Parish Council of Faj\~ao-Vidual.   We warmly thank George Smoot for its pioneering support and guidance without whom the installation of the GEM-P radiotelescope that later turned into a space radar would not be possible. The GEM project has been supported in part by the US Department of Energy, by the US National Science Foundation, by the Progetto Italia Antartide, by INPE-Brazil, and by Colciencias of Colombia. DB would like to warmly thank the Starlight Foundation and Instituto de Astrof\'{\i}sica de Canarias for the hospitality and excellent meeting.

\end{document}